MDPI

*Article*

# Periods of Outbursts and Standstills and Variations in Parameters of Two Z Cam Stars: Z Cam and AT Cnc

Daniela Boneva *, Krasimira Yankova and Denislav Rusev

Space Research and Technology Institute, Bulgarian Academy of Sciences, 1000 Sofia, Bulgaria;
f7@space.bas.bg (K.Y.); denislav.rusev@gmail.com (D.R.)
* Correspondence: danvasan@space.bas.bg

**Abstract:** We present our results on two Z Cam stars: Z Cam and AT Cnc. We apply observational data for the periods that cover the states of outbursts and standstills, which are typical for this type of object. We report an appearance of periodic oscillations in brightness during the standstill in AT Cnc, with small-amplitude variations of 0.03–0.04 mag and periodicity of $\approx$20–30 min. Based on the estimated dereddened color index $(B - V)_0$, we calculate the color temperature for both states of the two objects. During the transition from the outburst to the standstill state, Z Cam varies from bluer to redder, while AT Cnc stays redder in both states. We calculate some of the stars' parameters as the radii of the primary and secondary components and the orbital separation for both objects. We construct the profiles of the effective temperature in the discs of the two objects. Comparing the parameters of both systems, we see that Z Cam is definitely the hotter object and we conclude that it has a more active accretion disc.

## 1. Introduction

Cataclysmic variables (CVs) are binary stars, where the matter transfers from the main-sequence secondary to the white dwarf primary via the Roche lobe overflow [1,2].

Outbursts are the most obvious feature of these stars, which is also the source of their name [3]. These sudden eruptions can cause a rise in the magnitude of the corresponding star.

Four types of cataclysmic variable stars are known: novae, recurrent novae, dwarf novae, and novalike stars [1,4].

Novae/classical novae stars: for centuries, novae stars have meant that a new star appeared among the other stars and then disappeared. Today, it is known that novae stars are explosions of a neutron star, which is accreting matter from a main-sequence star. The difference for dwarf novae is that these outbursts of matter are much rarer [5,6].

Recurrent novae: they are the same as classical novae with the same mechanism of eruptions, but these eruptions have happened several times in recorded history [3].

Dwarf novae (DN): analyzed later in the text.

Novalike variables: they are almost the same objects as dwarf novae except for the fact that they do not exhibit outbursts [6].

Three subclasses of dwarf novae are known—U Gem stars, SU Uma stars, and Z Cam stars [4,7,8].

U Gem stars exhibit irregular quasi-periodic outbursts. The SU UMa type shows two kinds of outbursts: a short "normal" outburst, which last a few days and a long "superoutburst" that endures for approximately two weeks [5].

In this work, we examine one of these main subtypes, which is interesting due to its behavior—Z Cam stars. We focus on these objects because of their unusual properties seen in their light curves, which could be explained further by physical processes.

Z Cam stars exhibit two different states—outbursts and standstills [9]. The Z Cam stars that manifest a third, low state of quiescence [10] are not the case of our study in this paper. As dwarf novae stars, these binary systems also consist of a white dwarf and a main-sequence star [10].

Most of the Z Cam stars (Z Cam, RX And, WW Cet, HL CMa, AH Her, and SY Cnc) have similar minimal and maximal magnitudes and similar orbital periods [11].

Compared to many other dwarf novae, according to [10], Z Cam stars have longer orbital periods, higher mass transfer rates, and lower amplitudes of the outbursts.

They are studied mostly for their sequence of standstills and outbursts. Different objects experience different periods of standstills, which range from a few weeks to more than a year [7].

It is known that in CVs, an accretion disc is formed around the white dwarf. In dwarf novae during outbursts, the disc emission dominates the light from the system and the spectrum by the disc produces a blue continuum and broad emission lines of hydrogen, helium, and calcium [1].

A typical spectrum of Z Cam stars mainly consists of absorption lines [12].

All these features and more different aspects can be seen in many Z Cam stars studied by the authors. The three of these objects that were most paid attention are listed here.

A typical member of Z Cam stars is SY Cnc (Cancer), which is one of the brightest of its class. SY Cnc has a maximum magnitude of 11 [11]. It exhibits regular outbursts every 27 days [13], which is similar to the other objects of its class. According to [14], the secondary star of the binary could be more massive, with a mass of $\sim$1 $M_\odot$, which is untypical for this class of objects.

The Z Cam star AY Psc (Piscium) is the only eclipsing system from the Z Cam stars [15]. Its orbital period is approximately 5.22 h [16], with massive white dwarf mass estimated to be from 0.90 $M_\odot$ [17] to $\sim$1.31 $M_\odot$ and the mass of the secondary to be $\sim$0.59 $M_\odot$ [18]. The white dwarf's temperature reached values of 27,600–30,000 K [17].

IW And is a subtype of Z Cam stars. Its typical feature is a sequence of standstills interrupted by brightening. These stars are also known as anomalous Z Cam stars [19]. The mass of the white dwarf is estimated to be 0.75 $M_\odot$ and the effective temperature is 25,000 K [10]. The distance to IW And is 700 pc [10]. It is seen that IW And possesses spectra similar to usual dwarf novae—a strong Balmer emission at quiescence and absorption at outburst [10].

In this paper, we study in detail two Z Cam stars—Z Cam and AT Cnc. Their characteristics are given below.

The first star, Z Cam (Camelopardalis), is the prototype star for this class of objects. It has an orbital period of 0.289 days and an average outburst interval of 23 days [20].

Z Cam has an M-type donor star, which is much cooler than previously expected and its effective temperature is $\approx$3575 K [21]. According to the study in [12], the radius of the white dwarf is $5.4 \times 10^8$ cm. The radius of the outer part of the accretion disc is $5.4 \times 10^{10}$ cm [8]. Z Cam is one of the brightest of its type [22]. Its apparent magnitude is in the range of 10.0–14.5 in the V band [11]. The distance to it is estimated to be from 200 pc [23] to 390 pc [24]. The accretion rate of Z Cam star is on average $3 \times 10^{-9}$ $M_\odot \text{yr}^{-1}$ [9]. One typical thing about this object is that a sequence of outbursts is followed by a standstill [25]. As the standstills' periods are usually known to stay in a quiescence, in Z Cam, this period between outbursts is mostly observed to be in the high state. On the other hand, another interesting feature of the prototype star, and also for other Z Cams, is that during the standstill, the star is quite active [25].

The second object AT Cnc (Cancer) has an orbital period of 0.2011($\pm$0.0006) days and the mass of the primary star is 0.9($\pm$0.5) $M_\odot$ [26]. It was found by Bond and Tifft in 1974 [27] that there are shallow, broad absorption lines in the AT Cnc spectrum. The apparent magnitude of AT Cnc varies between 12.5 and 15.5 (AAVSO data). The distance to the object is estimated as 460 pc [28]. The outbursts and standstills of AT Cnc usually last a few weeks but sometimes the star has very long standstills—one in approximately half of a year [29]. It demonstrated that after the eruption, the mass transfer rate decreases substantially [28]. Kozhevnikov suggests that AT Cnc is the first star of Z Cams type in which superhumps are detected. These superhumps are observed during the long standstill of AT Cnc. They are typical of another type of cataclysmic variables—SU UMa stars.

The system parameters' values of Z Cam and AT Cnc are listed in Section 3.

The observational data for these two objects, obtained from AAVSO (American Association of Variable Star Observers), are reported in Section 2. In Section 3, we derive the color indices and temperatures at maximum brightness. Some system parameters as radius of the primary and secondary components, binary separation and the profiles of the discs' effective temperatures are calculated in Section 4. In Sections 4 and 5, we compare the obtained results for both objects.

## 2. Observational Data and Observational Effects

In this paper, we apply observational data, obtained from AAVSO. For Z Cam, we consider an observational period of time: 2458100–2459650 JD (Julian dates) (3 September 2017 to 16 November 2022—Calendar dates) in BVRI bands (see Figure 1). The data for AT Cnc are presented in BVR bands and the observational time is in the range: 2458100–2458220 JD (12 December 2017 to 11 April 2018) (see Figure 2).

The long-period light curve of Z Cam (Figure 1a) clearly shows a sequence of alternating events: an outburst period as a pre-standstill, a standstill period and again an outburst period as after-standstill, followed by a new standstill. During the whole time of the considered outburst period, 2458100–2458400, the amplitude variations reach ~3–3.5 mag in a frame of 1–2 weeks (Figure 1b), with a maximum value of 10.592 mag in V and 10.499 mag in B. The standstill state depicts small-amplitude variations in a brightness of ~0.2 magnitude for whole the time: 2458400–2459400. The star still stays in its high state with maximum magnitudes: 11.552 in V; 11.559 in B; 11.385 in R; 11.192 in I. The typical 1σ error's range is 0.04–0.06 mag in B and V. The standstill part of the light curve is shown in Figure 1c.

The light curve of AT Cnc includes one period of outbursts (2458090–2458145 JD) followed by a standstill period (2458150–2458220 JD) after it (Figure 2). The maximum brightness in the time of outbursts (Figure 2a) reaches 13.086–12.615 in V. The maximum magnitude in B is 13.124 and it is 12.936 in R. The typical 1σ error's range is 0.05–0.06 mag in B and V.

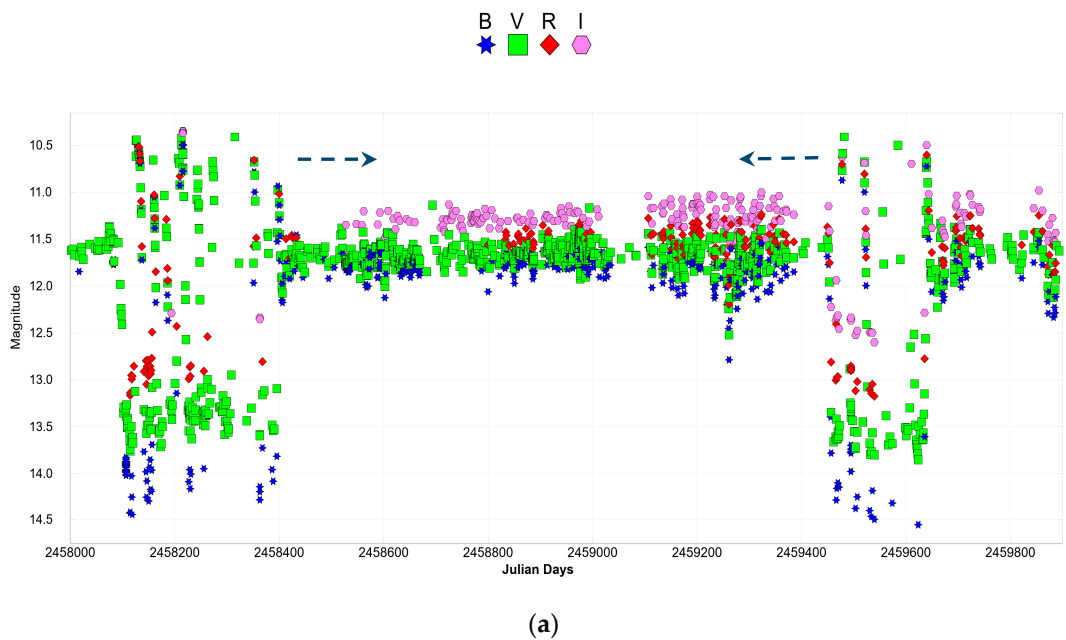

**(a)**

**Figure 1.** *Cont.*

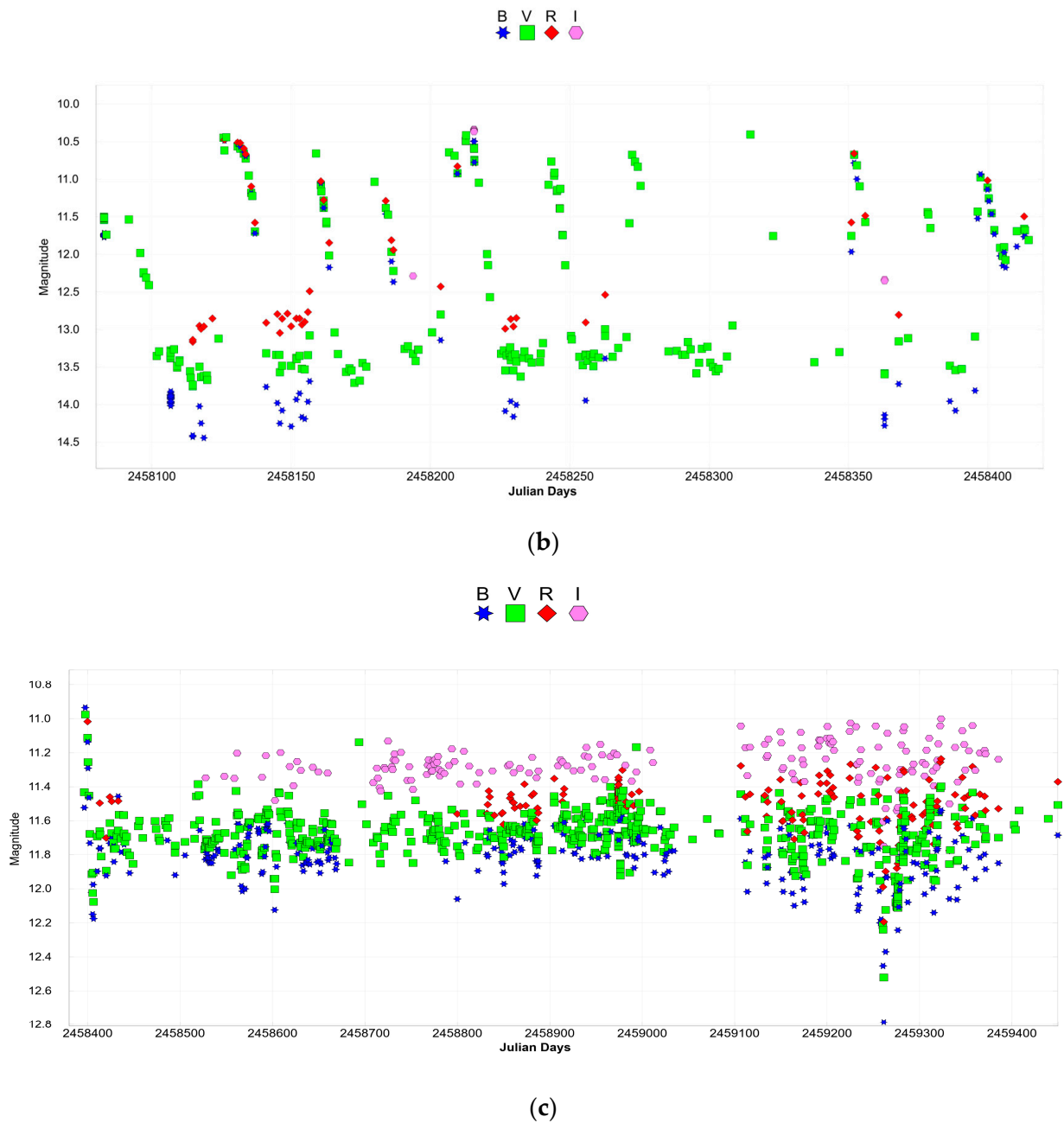

**Figure 1.** Light curves of Z Cam. The long period observations in BVRI bands (2458100–2459900 JD) (**a**) show the transitions between the large-amplitude variations in brightness (outbursts, 2458100–2458400) and the periods of quiescence in the standstills (marked with dark blue dashed arrows). The two states are separated: (**b**) for the outbursts period; (**c**) for the standstill period. The data are taken from AAVSO (Z Cam's most contributed observers codes: HKEB, SGEA, NOT, RFDA, LPAC, DGSA).

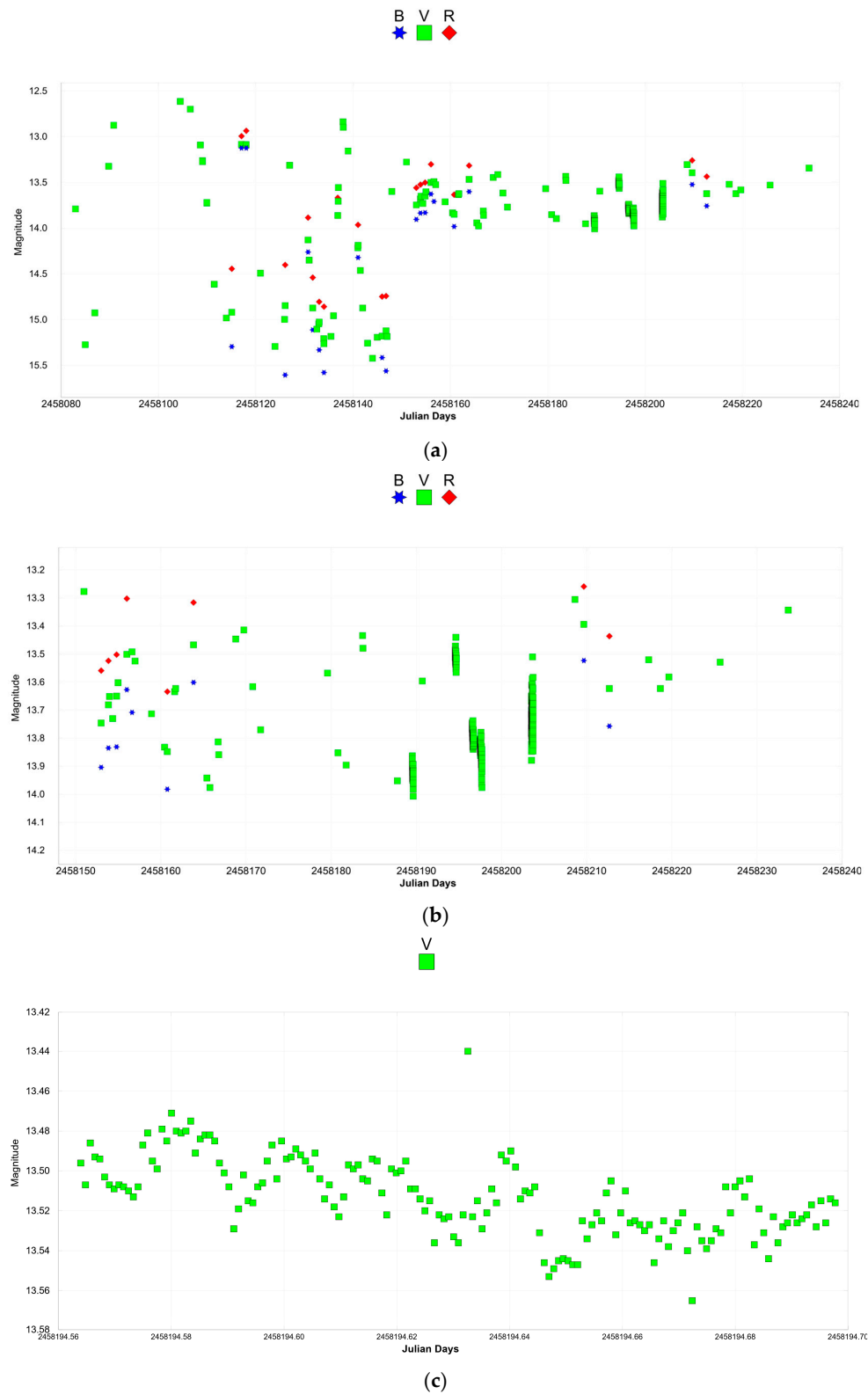

**Figure 2.** Light curves of AT Cnc. The long period observations in BVRI bands (2458080–2458240 JD) show the transitions between the large-amplitude variations in brightness (outbursts) and the period of quiescence in the high state (standstill) (**a**). The standstill state is shown in (**b**), and the zoomed oscillations in brightness during this period in (**c**). The data are taken from AAVSO (AT Cnc's most contributed observers' codes: CMJA, MZK, SGEA, PSD).

During the end of the considered standstill period of AT Cnc (2458150–2458220 JD), a trend of decreasing average brightness in V with ~0.07 magnitude is observed (Figure 2b, 2458197.50–2458197.70 JD). The brightness during the time 2458194.56–2458194.70 JD of the standstill period (Figure 2c) is fluctuating with amplitude variations of 0.03–0.04 mag and periodicity of 20–30 min in a frame of ≈3 h. Such short-term, small-amplitude brightness variations could be associated with flickering activity, which is usual for CVs [2,30,31] and we suppose they are seen in the light curve of Figure 2c.

Other short-period, low-magnitude brightness variations that typically occur in CVs are known as superhumps [32]. Superhumps' periodicity is a few percent longer (or shorter for the negative superhumps) than the binary period and they can be observed during the outbursts state of the objects [33]. The estimated periodicity of the brightness variations in AT Cnc is ≈10 times shorter (during the standstills) than the orbital period (4.82 h, see the table in page 8). Throughout the considered observational period of this object, we do not identify the appearance of superhumps activity.

## 3. Color Indices and Color Temperatures

In this study, the observations in BVRI filters enable estimating the color indices of Z Cam and AT Cnc. By analyzing the data, we obtain the indices at maximum brightness for both the periods of outbursts and standstill separately in the B and V filters. According to data availability, the values are calculated for different nights. The $B - V$ indices at a maximum brightness of Z Cam vary from slightly negative to positive values during both periods—the pre-standstill outbursts and standstill periods. This is an indication that the object becomes bluer and hotter during the outbursts and it is cooling while reddening in the standstill times. The obtained $B - V$ for AT Cnc also shows lower, but positive values during the outburst period, and the object remains red.

To be more precise in our estimations, we obtain the dereddened color index $(B - V)_0$. We use the color excess value $E(B - V) = 0.02 \pm 0.01$ for Z Cam (by Klare et al., 1982 in [11]). For AT Cnc, we obtain $E(B - V) = 0.0006 \pm 0.0001$, using the standard extinction law and $E(B - V) = Av/Rv$, where parameter $Rv = 3.1$ [34]. The averaged extinction $Av$ is derived by the known formula $m_v - Mv = 5 \log_{10}(d) - 5 + Av$, where d is the distance to the object (for AT Cnc, we use 460 pc [28]), m is the apparent magnitude and M is the absolute magnitude value in V. Then, we calculate $(B - V)_0$ for the two objects by $E(B - V) = (B - V) - (B - V)_0$.

Further, using the dereddened color index $(B - V)_0$, we could calculate the color temperature $T(B - V)_0$ for the two objects during their maximum brightness in both states. The formula of Ballesteros, (2012) [35] is well fitted for the whole spectrum and is appropriate to apply here:

$$(Tcol)_0 = 4600K \times \left[ \frac{1}{(0.92(B - V)_0 + 1.7)} + \frac{1}{(0.92(B - V)_0 + 0.62)} \right] \tag{1}$$

where $(Tcol)_0$ from Equation (1) corresponds to $T(B - V)_0$. The resulting color temperatures confirm the variations in values and it becomes highest at maximum brightness during the outburst's period; for both objects: $T(B - V)_0 = 11{,}800 \pm 900$ K for Z Cam and $T(B - V)_0 = 9700 \pm 500$ K for AT Cnc. The calculated standstill temperatures at maximum brightness are $T(B - V)_0 = 9800 \pm 700$ K and $8800 \pm 400$ K, respectively.

The observed color indices $B - V$, the dereddened color index $(B - V)_0$ and the color temperatures for both objects are given in Table 1.

**Table 1.** Color index and color temperatures, values at maximum brightness of the outbursts and standstill states.

| Object/ Values | Outbursts (Pre-Standstill) | | | Standstill | | |
|---|---|---|---|---|---|---|
| | B − V_max | (B − V)_0_max | (Tcol)_0_max | B − V_max | (B − V)_0_max | (Tcol)_0_max |
| | JD 2458215.62803 | | | JD 2458975.6455–2458975.64692 | | |
| Z Cam | $-0.09 \pm 0.03$ | $-0.113 \pm 0.04$ | $11{,}800 \pm 900$ | $0.04 \pm 0.02$ | $0.024 \pm 0.02$ | $9800 \pm 700$ K |
| | JD 2458117.06694 | | | JD 2458209.66426–2458209.66597 | | |
| AT Cnc | $0.04 \pm 0.03$ | $0.037 \pm 0.03$ | $9700 \pm 500$ K | $0.13 \pm 0.03$ | $0.126 \pm 0.03$ | $8800 \pm 400$ K |

Comparing the results, we see that Z Cam is the hotter object and it is valid for both states of outburst and standstill behavior.

## 4. System Parameters, Calculation and Comparison

### 4.1. Radii and Orbital Separations

In this section, we present some of the system parameters of Z Cam and AT Cnc. We aim to compare the values of both objects, as part of them are calculated in this paper.

We have the values of the primary, white dwarf stars' masses ($M_1$) for both objects from the literature (see Table 2 for the values). To calculate the radius of the white dwarf primary $R_1$, the Eggleton's mass–radius relation is convenient to use, since it is applicable for the mass range $0 < M_{1,2} < M_{Ch}$ [36,37]:

$$\frac{R_1}{R_\odot} = 0.0114 \left[ \left( \frac{M_1}{M_{Ch}} \right)^{-2/3} - \left( \frac{M_1}{M_{Ch}} \right)^{2/3} \right]^{1/2} \times \left[ 1 + 3.5 \left( \frac{M_1}{M_p} \right)^{-2/3} + \left( \frac{M_1}{M_p} \right)^{-1} \right]^{-2/3} \tag{2}$$

where $M_{Ch} = 1.4\ M_\odot$ is the Chandrasekhar mass; $M_p = 0.00057\ M_\odot$ is a constant. The calculations for the radius of the primary for both objects give the results:

$$R_1 = (3.11 \pm 0.006) \times 10^8\ \text{cm} = 0.004 \pm 0.002\ R_\odot\ \text{for Z Cam;}$$

$$R_1 = (3.96 \pm 0.050) \times 10^8\ \text{cm} = 0.005 \pm 0.003\ R_\odot\ \text{for AT Cnc.}$$

The calculated radius of the Z Cam's primary component differs from the value obtained in [7] (see target details) by a factor of Section 2.3.

The orbital separation "a" can be calculated by applying Kepler's 3rd law expression:

$$a = \left( \frac{G(M_1 + M_2)M_\odot \text{P}_{orb}^2}{4\pi^2} \right)^{\frac{1}{3}} \tag{3}$$

where $M_2$ is the mass of the secondary component (Z Cam [8], AT Cnc [29]), Porb is the orbital period (Z Cam [11], AT Cnc [26]), and G is the gravitational constant.

The obtained values of the orbital separation are as follows:

$$a\ (\text{Z Cam}) = (1.51 \pm 0.01) \times 10^{11}\ \text{cm} = 2.18 \pm 0.021\ R_\odot$$

$$a\ (\text{AT Cnc}) = (1.09 \pm 0.02) \times 10^{11}\ \text{cm} = 1.58 \pm 0.03\ R_\odot$$

In the case of the CVs, they are at the evolutionary point of time when mass transfer has already started. This involves the dwarf novae to which our objects in the current study belong. Then, we could accept the secondary star should have fulfilled its Roche lobe, and its radius $R_2$ is approximately equal to the Roche lobe $R_L$, and $R_2 \approx R_L$.

Further, to calculate the radius $R_2$ of the secondary components of the two objects, the equation of Eggleton (1983) [36] is applicable. It gives a high accuracy (~1%) of the obtained values and q could be in the range $0 < q < \infty$:

$$\frac{R_2}{a} = \frac{0.49q^{\frac{2}{3}}}{0.6q^{\frac{2}{3}} + ln\left(1 + q^{\frac{1}{3}}\right)} \tag{4}$$

We obtain for the radii:

$$R_2 \ (Z \ Cam) = (5.26 \pm 0.001) \times 10^{10} \ cm = 0.762 \pm 0.002 \ R_\odot$$

$$R_2 \ (AT \ Cnc) = (3.12 \pm 0.004) \times 10^{10} \ cm = 0.451 \pm 0.001 \ R_\odot$$

The set of parameters' values are given in Table 2.

**Table 2.** System parameters of Z Cam and AT Cnc, as follows: Porb—orbital period; M1—mass of the primary component; M2—mass of the secondary component; q—mass ratio; Teff—effective temperature of the white dwarf; R1—radius of the primary; R2—radius of the secondary; a—binary separation.

| Object/ Parameter | Porb [days] | $M_{wd}$ [$M_\odot$] | $M_2$ [$M_\odot$] | q | Teff [K] (of WD) | R1 [$R_\odot$] | A [$R_\odot$] | R2 [$R_\odot$] |
|---|---|---|---|---|---|---|---|---|
| Z Cam | 0.289 [11] | 0.99 ± 0.15 [12] | 0.70 ± 0.02 [8] | 0.71 ± 0.10 [12] | 57,000 [38] 150,000 [12] | 0.004 ± 0.002 [tp] | 2.18 ± 0.021 [tp] | 0.762 ± 0.002 [tp] |
| AT Cnc | 0.201 ± 0.0006 [26] | 0.9 ± 0.5 [26,29] | 0.47 ± 0.05 [29] | 0.52 ± 0.08 [29] | 13,500 ± 100 [39] 40,000 ± 170 [40] | 0.005 ± 0.003 [tp] | 1.58 ± 0.031 [tp] | 0.451 ± 0.001 [tp] |

[tp] This paper calculations.

### 4.2. Effective Temperature Profiles of the Accretion Discs

Observations of AT Cnc show the presence of flickering, which is an indication of activity in the disc flow. That is why, in this section, we will use the formula of Pringle 1981 [41] to demonstrate the development of the effective temperature in the discs of the two objects.

$$T_{eff} = T_{wd}\left(\frac{R_{wd}}{r}\right)^{\frac{3}{4}}\left(1 - \left(\frac{R_{wd}}{r}\right)^{\frac{1}{2}}\right)^{\frac{1}{4}} \tag{5}$$

The surface temperature of young white dwarfs reaches over $2 \times 10^5$ K, while the coldest ones are approximately $5 \times 10^3$ K. The results for the surface temperatures of accretors in the selected objects provided by different authors (see Table 2) significantly differ by orders beyond the permissible error (~200 K).

Figure 3a,b illustrate how the disc temperature profile changes from a low to a high possible limit for the surface temperature of the accretor in each object. The effective temperatures are plotted against the discs' radii (Rd), with Rd measured in units of white dwarf radius $R_1$. We will consider the higher values as valid, because as we mentioned earlier in the text, these are heavy dwarfs, and as we know, they are with more complex chemical composition and internal structure. The selected objects are C/O dwarfs by the mass scale (He-dwarfs [0.01–0.45] $M_\odot$; hybrids dwarfs; C/O dwarfs [0.5–1.05] $M_\odot$ and Mg/Ne/O dwarfs $\geq 1.05$ $M_\odot$) and they are near the upper limit of this interval $\simeq 1.05$ $M_\odot$ [42]. Therefore, the cooling process is slowed down by crystallization [43,44] and the dwarfs' surfaces still remain hot. As a result, high-temperature profiles of the disc should be considered in both cases.

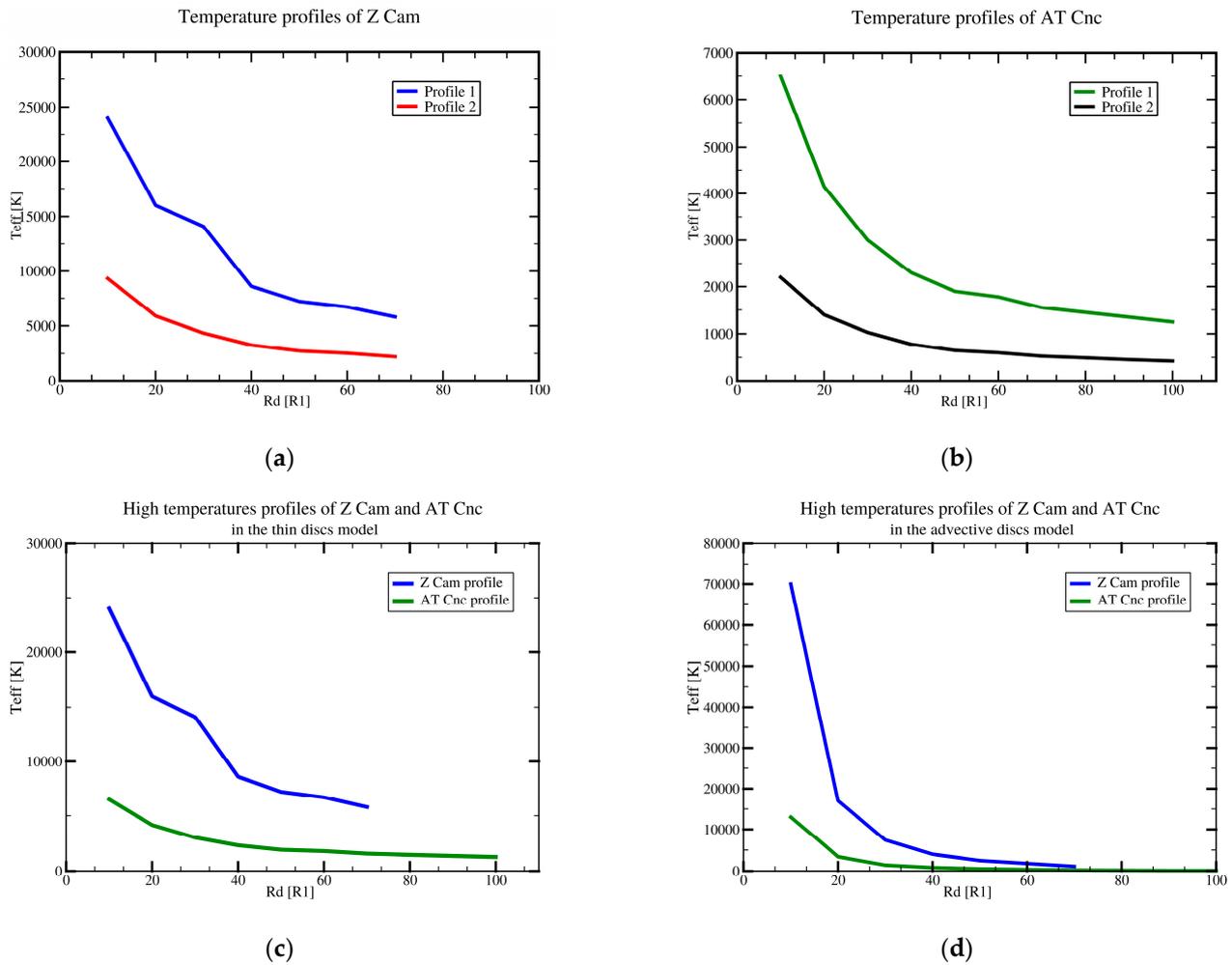

**Figure 3.** The effective temperatures profiles of Z Cam and AT Cnc against the accretion disc's radius. Two profiles are presented separately in (**a**,**b**). A comparison of the two objects profiles is seen in (**c**) for the thin model and (**d**) for the advective model.

The effective temperature profiles comparison in the discs of Z Cam and AT Cnc is given in Figure 3.

The white dwarf with a larger radius should be colder on its surface. The difference of 8500 km in the radii contributes to nearly three times lower temperature of the white dwarf surface for AT Cnc. The hot accretor is more compact and the disc radius Rd is significantly smaller, as can be seen in the next ratio:

$$\text{Rd (Z Cam)/Rd (AT Cnc)} = 0.55 \pm 0.02,$$

where this suggests the accelerated action of processes in the disc, an increased activity in the stream and, accordingly, a higher accretion efficiency, which additionally supports the high temperature in the disc's flow of the Z Cam.

We can confirm this conclusion with two other profile distributions calculated in our model (presented in [45–48]), which is more useful for hot compact accretors than the thin disc model.

We apply one-temperature model of an accretion disc and the disc develops an advection. With this model, a non-stationary and non-axisymmetric accretion flows are investigated, but the advection remains in the non-dominant regime/mode. The model's results allow us to obtain the 2D structure of the disc [49] and then Teff in terms of this solution is:

$$T_{eff} = T_0 \left( \frac{x}{R_d} \right)^{\frac{1}{4}} \left( f_1 f_4 x (4x)^3 \right)^{\frac{1}{4}}$$ (6)

Here, $T_0$ is the temperature initial's value at the outer edge of the disc Rd; x = r/Rd, $f_1(x)$ = equatorial density and $f_4(x)$ = disc half-thickness, correspondingly:

$$\text{for Z Cam :} \quad f_1(x) \approx 23(1-x)x^{-7.5}$$ (7)

$$\text{for AT Cnc :} \quad f_1(x) \approx 41(1-x)x^{-7.5}$$ (8)

$$\text{for both objects :} \quad f_4(x) \approx 0.1x^{-6}$$ (9)

Then the profiles have the form:

$$\text{for Z Cam :} \quad T_{eff} \approx 5 \left( \frac{10}{x} \right)^2 (4.7(1-x))^{\frac{1}{4}}$$ (10)

$$\text{for AT Cnc :} \quad T_{eff} \approx \left( \frac{10}{x} \right)^2 (4.1(1-x))^{\frac{1}{4}}$$ (11)

The difference in the temperature profiles of the two discs is even more clearly expressed in Figure 3d. As seen from the results, the advective model much better describes the correlation between the disc's active nature and the degree of compactness of the accretor.

## 5. Discussion

With our results of Z Cam and AT Cnc, we contributed to enrich the knowledge of the stars' color index and temperature, and also their parameters like the radii of the primary and secondary components, and the orbital separation for both objects. By the calculations of the two Z Cam objects' radii, we find that the values are typical for their class. Their white dwarfs are quite massive and we see a difference of approximately 0.23 $M_\odot$ in masses of the companion stars. Comparing the radii of the secondary components in Z Cam and AT Cnc, we see that in the first object, R2 is almost two times larger than the R2 of the second object. In the same time, the distance between the components in Z Cam is approximately 2.18 $R_\odot$ and it is bigger than the obtained for AT Cnc components ≈ 1.58 $R_\odot$. This could result in further calculations of the mass transfer between the components, which is partially responsible for the outbursts and flickering activity in the Z Cam's objects.

As we noted earlier, the disc emission dominates the light from the system during outbursts of the dwarf novae objects. This is an indication that an outburst may occur in any of the active areas of the disc. The high effective temperatures (as our results show in Figure 3c,d) during the standstill state keep an activity in the disc stable. Then, an excess in energy is released in the disc's flow, which feeds the local processes for the specific active zone and could provoke a new cycle of outbursts in this disc's zone. As expected, the system of the more active disc around the more compact accretor shows significantly more powerful outbursts ≈2 mag (Figure 1) and ≈0.5 mag (Figure 2).

The flickering activity appears in the standstill state of the cooler object (AT Cnc), and we suggest that their source could be fluctuations in the disc due to the developed instabilities. In Z Cam, this effect is suppressed by the high temperature background in the disc flow.

## 6. Conclusions

From our analysis of two Z Cam stars, Z Cam and AT Cnc, we can present the following conclusion: both objects are passing the periods of standstills and outbursts during the selected observational times, which makes them typical for their class.

Based on the observational data, we estimated the color indices and color temperatures for the two states separately at their maximum brightness. From the outbursts to the standstills, the values for Z Cam are in the intervals: $-0.113 < (B - V)_0 < 0.024$; $11,800$ [K] $> T(B - V)_0 > 9800$ [K]; and for AT Cnc: $0.037 < (B - V)_0 < 0.126$; $9700$ [K] $> T(B - V)_0 > 8800$ [K].

Comparing the values for the two stars, it turns out that AT Cnc is a much cooler object than Z Cam.

We found the appearance of small-amplitude brightness variations of 0.03–0.04 mag in V (or flickering activity) during the standstill period of AT Cnc.

We calculated the radii of the primary and secondary components, and the orbital separation for both objects.

We constructed the profiles of the effective temperature in the accretion discs of the two objects. This confirms that Z Cam is definitely the hotter object and its disc is more active during both states.

**Author Contributions:** Conceptualization, D.B. and K.Y.; methodology, D.B. and K.Y.; software, D.B.; formal analysis, D.B., K.Y. and D.R.; resources, D.B., K.Y. and D.R.; data selection, D.B. and D.R.; writing—original draft preparation, D.B., K.Y. and D.R.; visualization, D.B. and K.Y. All authors have read and agreed to the published version of the manuscript.

**Funding:** This research received no external funding.

**Institutional Review Board Statement:** Not applicable.

**Informed Consent Statement:** Not applicable.

**Data Availability Statement:** Observational data are available at www.aavso.org. Accessed on 29 March 2024 and 29 April 2024.

**Acknowledgments:** The authors acknowledge with thanks the variable star observations from the AAVSO International Database contributed by observers worldwide and used in this research. The authors thank the anonymous referees for the useful comments that helped us improve this paper.

**Conflicts of Interest:** The authors declare no conflicts of interest.